\documentclass[twocolumn,tighten,twocolappendix]{aastex631}
\usepackage{graphicx}
\usepackage{amsmath}
\usepackage{amsfonts}
\usepackage{amssymb}
\usepackage{natbib}

\begin{document} 
%\nolinenumbers
   \title{Photospheric Hot Spots at Solar Coronal Loop Footpoints Revealed by Hyperspectral Imaging Observations}

\correspondingauthor{L. P. Chitta}
\email{chitta@mps.mpg.de}

\author[0000-0002-9270-6785]{L. P. Chitta}
\affiliation{Max-Planck-Institut f\"ur Sonnensystemforschung, Justus-von-Liebig-Weg 3, 37077 G\"ottingen, Germany}

\author{M. van Noort}
\affiliation{Max-Planck-Institut f\"ur Sonnensystemforschung, Justus-von-Liebig-Weg 3, 37077 G\"ottingen, Germany}

\author[0000-0003-3490-6532]{H. N. Smitha}
\affiliation{Max-Planck-Institut f\"ur Sonnensystemforschung, Justus-von-Liebig-Weg 3, 37077 G\"ottingen, Germany}

\author[0000-0003-3621-6690]{E. R. Priest}
\affiliation{School of Mathematics and Statistics, University of St Andrews, St Andrews, KY16 9SS, UK}

\author[0000-0003-2088-028X]{L. H. M. Rouppe van der Voort}
\affiliation{Institute of Theoretical Astrophysics, University of Oslo, PO Box 1029 Blindern, 0315 Oslo, Norway}
\affiliation{Rosseland Centre for Solar Physics, University of Oslo, PO Box 1029 Blindern, 0315 Oslo, Norway}

\begin{abstract}
Poynting flux generated by random shuffling of photospheric magnetic footpoints is transferred through the upper atmosphere of the Sun where the plasma is heated to over 1\,MK in the corona. High spatiotemporal resolution observations of the lower atmosphere at the base of coronal magnetic loops are crucial to better understand the nature of the footpoint dynamics and the details of magnetic processes that eventually channel energy into the corona. Here we report high spatial resolution ($\sim$0.1\arcsec) and cadence (1.33\,s) hyperspectral imaging of the solar H$\alpha$ line, acquired by the Microlensed Hyperspectral Imager prototype installed at the Swedish 1-m Solar Telescope, that reveal photospheric hot spots at the base of solar coronal loops. These hot spots manifest themselves as H$\alpha$ wing enhancements, occurring on small spatial scales of $\sim$0.2\arcsec, and timescales of less than 100\,s. By assuming that the H$\alpha$ wings and the continuum form under the local thermodynamic equilibrium condition, we inverted the H$\alpha$ line profiles and found that the hot spots are compatible with a temperature increase of about 1000\,K above the ambient quiet-Sun temperature. The H$\alpha$ wing integrated Stokes\,$V/I$ maps indicate that hot spots are related to magnetic patches with field strengths comparable to or even stronger than the surrounding network elements. But they do not show the presence of parasitic polarity magnetic field that would support the interpretation that these hot spots are reconnection-driven Ellerman bombs. Therefore, we interpret these features as proxies of locations where convection-driven magnetic field intensification in the photosphere can lead to energy transfer into higher layers. We suggest that such hot spots at coronal loop footpoints may be indicative of the specific locations and onset of energy flux injection into the upper atmosphere. 
\end{abstract}

%
%-------------------------------------------------------------------
%
\section{Introduction\label{sec:intr}}
The processes that generate and sustain the million degree Kelvin hot solar corona continue to be a subject of active debate in the solar community. It is commonly accepted that coronal heating is driven by the magnetic field that has its roots in the photosphere. Random buffeting of the magnetic field by turbulent motions in the intergranular lanes of convective cells generates the energy flux. Current sheets \citep[e.g.,][]{1972ApJ...174..499P,2003Natur.425..692S}, magnetohydrodynamic (MHD) waves \citep[e.g.,][]{2007Sci...318.1574D,2020SSRv..216..140V}, and reconnection powered jets \citep[e.g.,][]{2018ApJ...862L..24P,2019ApJ...872...32S} are generally considered conduits of energy transfer through the solar atmosphere. Thus probing the dynamic coupling between turbulent convective motions and the magnetic field in the intergranular lanes near the solar surface offers a clearer picture of the onset and nature of the energy flux injection into the solar atmosphere. 

Observations with sub-arcsec spatial resolution are important to study the magnetic field dynamics confined to the narrow width of intergranular lanes ($\sim$100\,km). For decades now photospheric observations acquired with broad-band filters (e.g., white light, G-band) have been widely used to investigate the properties of small-scale bright points (manifestations of magnetic field patches) in both quiet-Sun and active regions \citep[e.g.,][]{1992Natur.359..307K,1995ApJ...454..531B}. Recording filtergrams at high-cadence ($\sim$5\,s) provides useful information such as the speeds of horizontal motions of bright points in the photosphere, which lie in the range of 0.5\,km\,s$^{-1}$ to 15\,km\,s$^{-1}$ \citep[e.g.,][]{2012ApJ...752...48C,2013A&A...549A.116J}.

\begin{figure*}
\begin{center}
\includegraphics[width=0.8\textwidth]{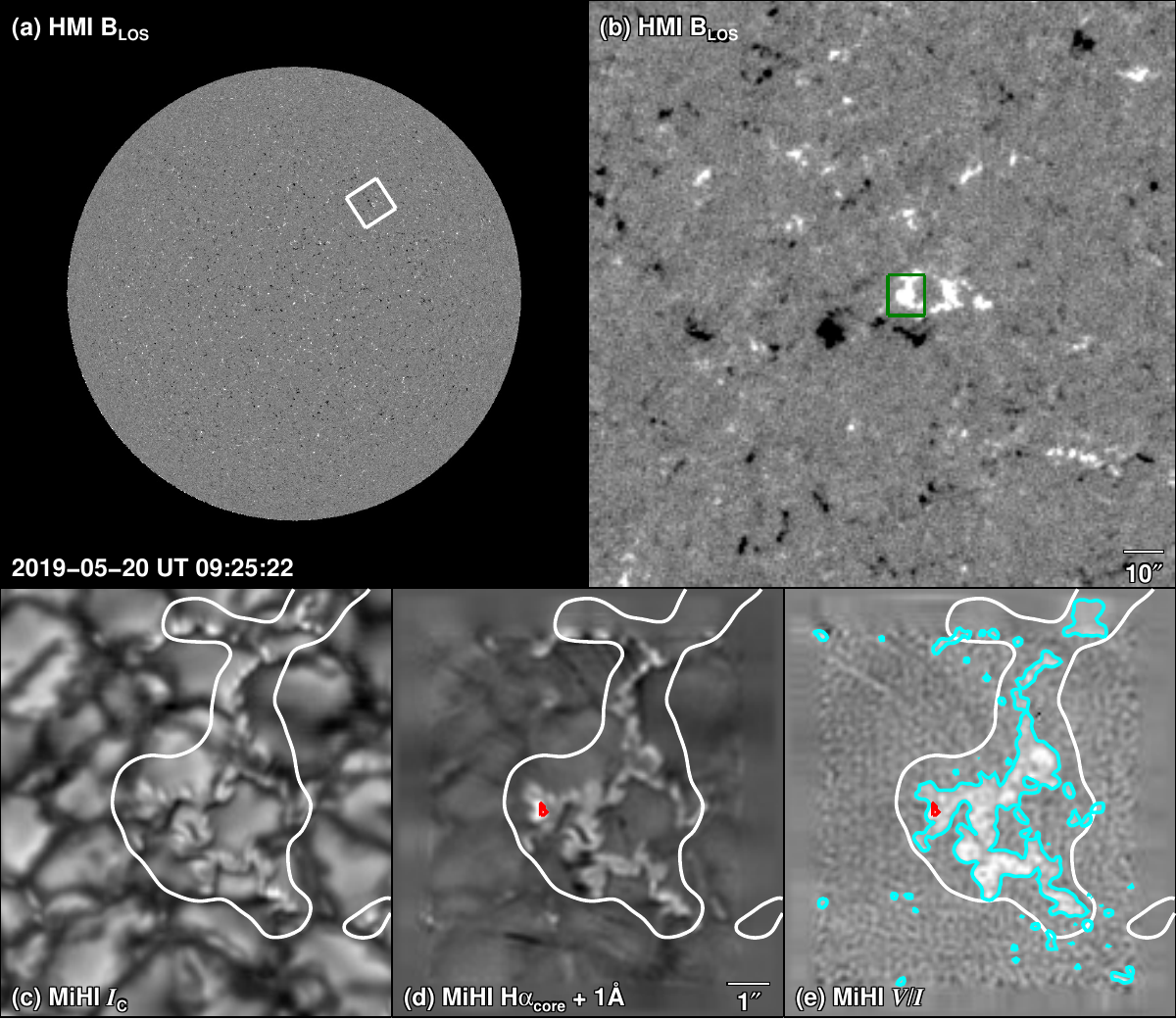}
\caption{Context for MiHI observations. Panel a: SDO/HMI full-disk map of the line-of-sight (LOS) component of the magnetic field, recorded on 2019 May 20 around UT\,09:25, saturated to $\pm75$\,G. Positive (negative) polarity magnetic field patches are represented by lighter (darker) shaded regions.  North is up. The white box, rotated by 57\textdegree\ (in the anticlockwise direction) about $(x, y)=(322.5\arcsec, 379.5\arcsec)$ in the Helioprojective-Cartesian coordinate system, outlines a magnetic bipole. Panel b: zoom-in map of the region outlined by the white box in panel a. The side length is about 150\arcsec. The green box (9.5\arcsec$\times$10.4\arcsec), covering the positive polarity patch, represents the field of view of MiHI hyperspectral imaging observations. Panels c and d: MiHI continuum intensity ($I_{\rm C}$) map around 6468\,\AA, and a map in the red wing of H$\alpha$. Panel e: Integrated MiHI Stokes $V/I$ map showing the network elements marked by cyan contours. The white contour in the lower panels encloses regions where the SDO/HMI LOS magnetic field component exceeds 75\,G. The red contour in panel d outlines a hot spot. For reference, a 1\arcsec\ scale is added in panel d. Movie associated with panel e, without contours, is available as an online animation. The movie is played back at 30\,frames\,s$^{-1}$, and the real-time animation duration is 11\,s. The movie has start and stop times from UT\,09:25:18 to UT\,10:30:56, at a cadence of 24\,s. See See Sects.\,\,\ref{sec:obs} and \ref{sec:det} for details.\label{fig:over}}
\end{center}
\end{figure*}

\begin{figure*}
\begin{center}
\includegraphics[width=0.7\textwidth]{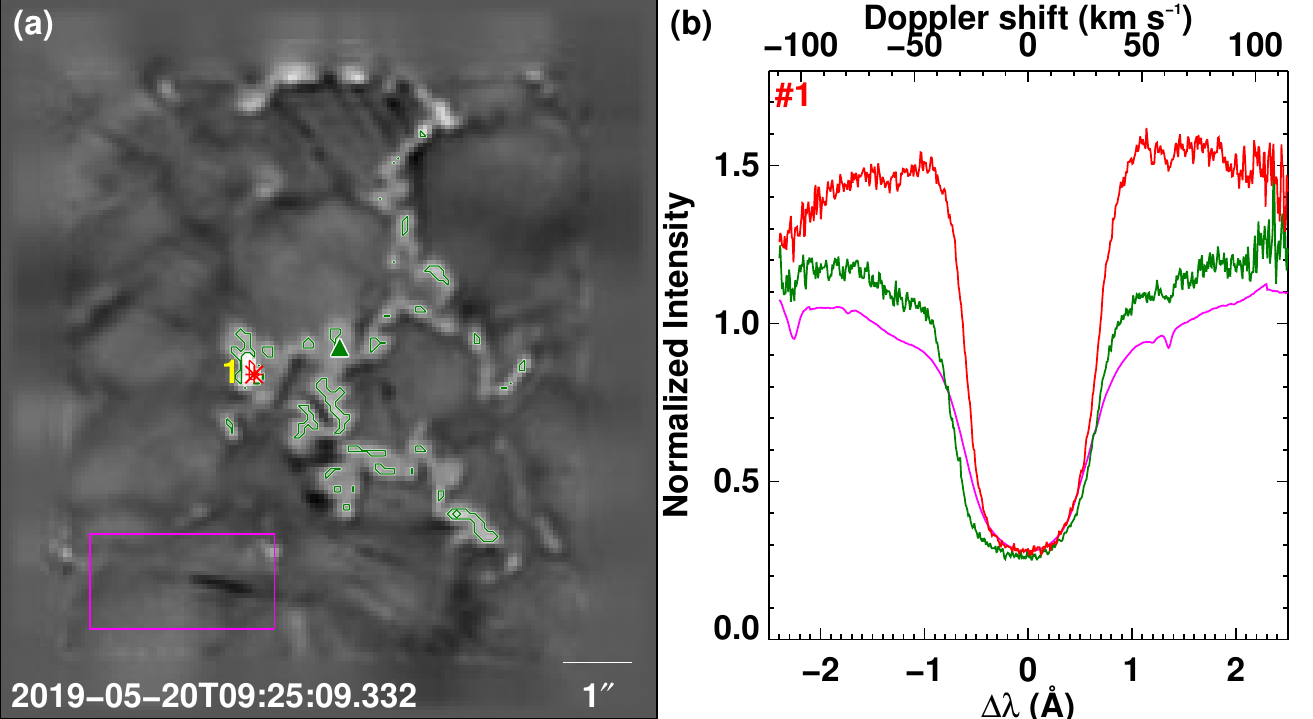}
\includegraphics[width=0.7\textwidth]{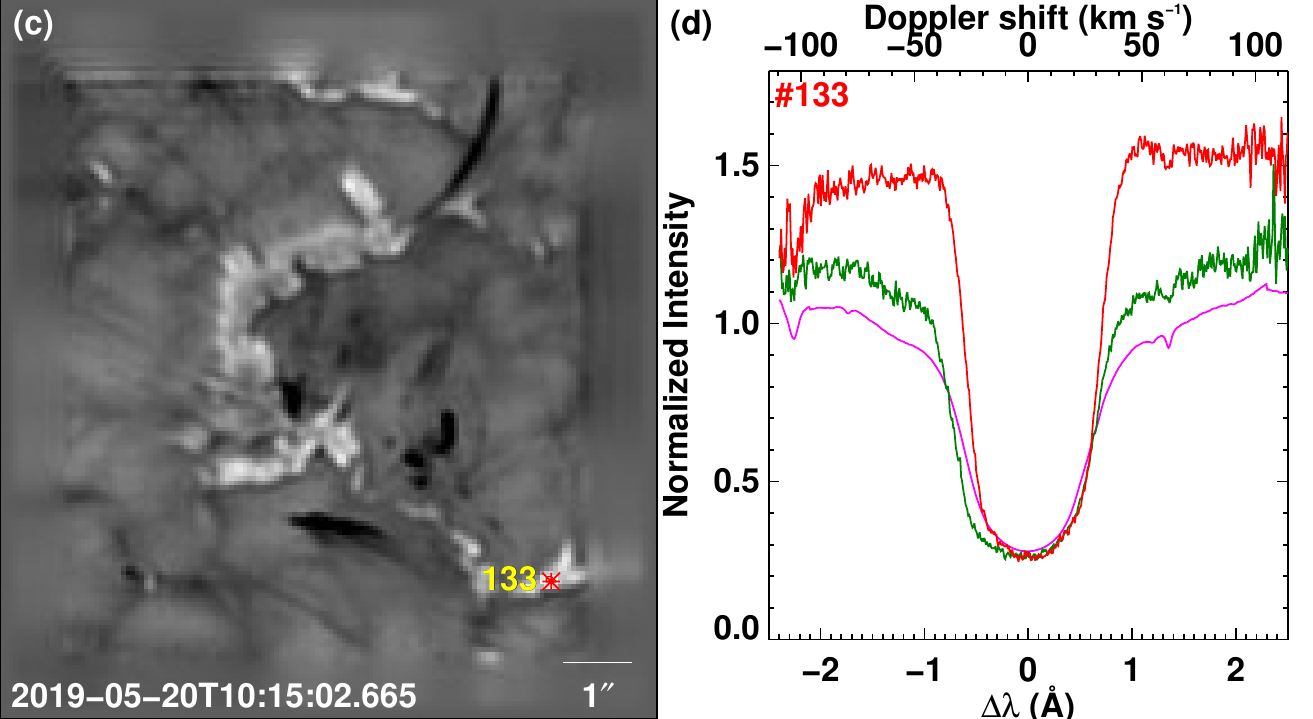}
\caption{Examples of hot spots. Panels a and c display H$\alpha$ red wing maps at two instances as indicated by the timestamps. The red stars (labeled with numbers) are the hot spots that we detected at those instances. In panel a the magenta box outlines a quiet-Sun patch away from the large brighter network patch spanning the center of the field of view. The green contours mark network magnetic elements, and one such feature is further identified with the green triangle. Panels b and d show H$\alpha$ spectral profiles from three types of features (red: hot spot region; green: network element; magenta: average profile from the quiet-Sun patch). In panel d the spectral profiles representing the network element (green triangle) and the quiet-Sun patch remain the same as plotted in panel b. An online animation in the format of panels c and d is available online, showing all the detected hot spots and their spectral profiles. The movie is played back at 30\,frames\,s$^{-1}$, and the real-time animation duration is 161\,s. The movie has start and stop times from UT\,09:25:09 to UT\,10:26:27. In the movie, the label of a given hot spot appears only once, that is in the frame or instance when its intensity peaks. Therefore, some of the contours in some frames that appear without any number are a part of a longer living hot spot, whose label would appear in a different frame. See Sect.\,\ref{sec:det} and Appendix\,\ref{app:detect} for details.\label{fig:events1}}
\end{center}
\end{figure*}

\begin{figure*}
\begin{center}
\includegraphics[width=0.7\textwidth]{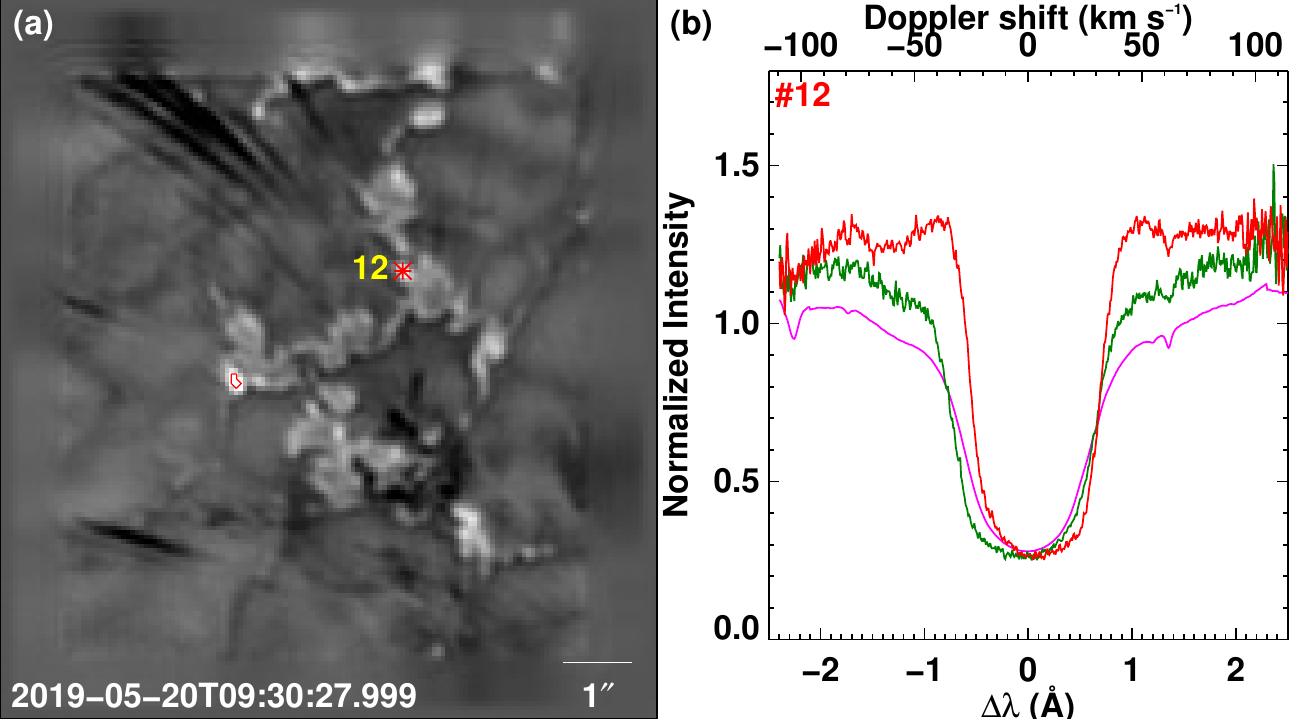}
\includegraphics[width=0.7\textwidth]{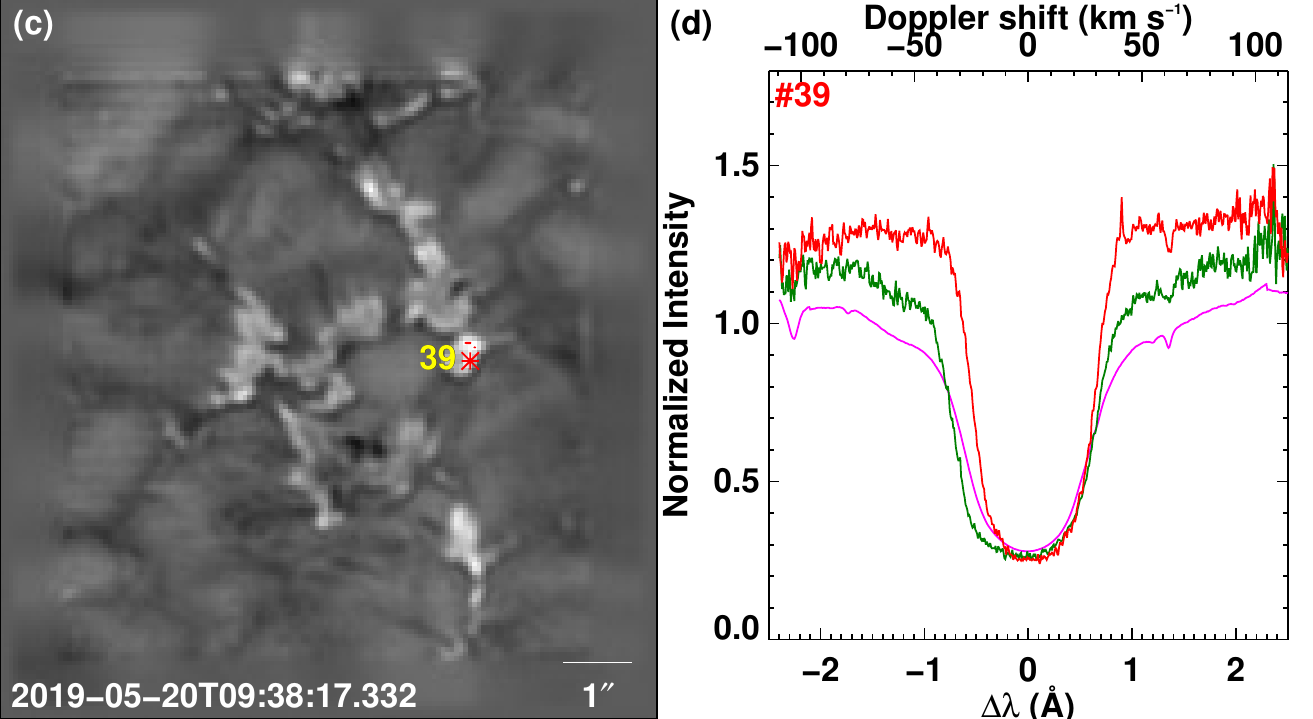}
\caption{Examples of hot spots. Same as Fig.\,\ref{fig:events1}, but showing additional examples.\label{fig:events2}}
\end{center}
\end{figure*}

Beyond filtergrams, spectroscopic and spectropolarimetric observations (obtained using Fabry-P\'{e}rot interferometers or slit spectrographs) offer a quantitative picture on the energetics of different magnetic processes and plasma dynamics in the photospheric layers and above. These include the observations fully resolved magnetic flux tubes in the quiet-Sun \citep[][]{2010ApJ...723L.164L}, different MHD wave phenomena \citep[e.g.,][]{2009Sci...323.1582J,2012ApJ...752L..12D,2017NatSR...743147S}, signatures of small-scale magnetic reconnection \citep[e.g.,][]{2016A&A...592A.100R,2017ApJS..229....4C,2020A&A...641L...5J}, and the existence of topological structures on granular-scales in the form of undulating fields \citep[][]{2023ApJ...955L..36C}.

Traditional spectroscopic observations suffer from a trade-off between spatial, temporal, and spectral resolutions. Due to the need to scan an extended field of view, slit spectrographs offer limited temporal information on rapidly evolving features, although they provide higher spatial resolution and spectral coverage. On the other hand, tunable filtergraphs based on the principle of Fabry-P\'{e}rot interferometry offer a limited spectral resolution and coverage, and thus limit the range of plasma dynamics that can be captured, despite their higher spatial and temporal resolution capabilities.

The recently designed prototype Microlensed Hyperspectral Imager \citep[MiHI;][]{2022A&A...668A.149V,2022A&A...668A.150V} is an integral field spectrograph, which overcomes the limitations of traditional spectroscopy by providing observations over a limited field of view and spectral range, but at a high cadence and spectral resolution \citep[][]{2023A&A...673A..11R}. In this study we used MiHI observations to investigate the nature of photospheric small-scale magnetic dynamics and signatures of energy flux injection at the base of coronal loops. 

\section{Observations\label{sec:obs}}
We used imaging spectropolarimetric observations of the H$\alpha$\,6563\,\AA\ spectra line recorded by a prototype Microlensed Hyperspectral Imager \citep[MiHI;][]{2022A&A...668A.149V,2022A&A...668A.150V}, installed at the Swedish 1-m Solar Telescope that is equipped with an adaptive optics system \citep[SST;][]{2003SPIE.4853..341S,2003SPIE.4853..370S,2019A&A...626A..55S}. The telescope was pointed at solar $(x, y)\sim(323\arcsec, 374\arcsec)$, with the cosine of the angle between the surface normal and the observer line-of-sight, $\mu=0.85$. The field of view has a roll angle of about 57\textdegree\ in the anticlockwise direction with respect to solar north. The observations were recorded for about 66\,minutes, starting at UT\,09:25 on May 20 2019.

The wavelength coverage was approximately $\pm$2.5\,\AA\ with respect to the nominal H$\alpha$ rest wavelength position of 6562.817\,\AA. The line profile was observed with a spectral sampling of 8.5\,m\AA\,pixel$^{-1}$ (corresponding to a Doppler shift of 0.4\,km\,s$^{-1}$). The observations also include a continuum point at a wavelength of approximately 6468\,\AA.

Reduction and calibration of MiHI data are detailed in \citet{2022A&A...668A.151V}, whereas the image restoration methodology using the multi-object multi-frame blind deconvolution technique are described in \citet{1994A&AS..107..243L,2005SoPh..228..191V}. The restored MiHI imaging spectropolarimetric dataset has an image scale of 0.065\arcsec\,pixel$^{-1}$ (1\arcsec\ corresponds to $\sim725$\,km on the Sun) and cadence of 1.33\,s, covering a field of approximately 7.6\arcsec$\times$8.5\arcsec. The MiHI data include images from a context wide-band imager, used in the image restoration of the spectropolarimetric data, covering a larger field of view, but those were not used in our analysis presented here. 

We complemented the MiHI data with the line-of-sight magnetic field observations acquired by the Helioseismic Magnetic Imager \citep[HMI;][]{2012SoPh..275..207S} onboard the Solar Dynamics Observatory \citep[SDO;][]{2012SoPh..275....3P}. These HMI data have a cadence of 45\,s and image scale of about 0.5\arcsec\,pixel$^{-1}$. For upper atmospheric coronal context we have used extreme ultraviolet (EUV) level-1 images from the Atmospheric Imaging Assembly \citep[AIA;][]{2012SoPh..275...17L} onboard the SDO. In particular, we considered images from the 193\,\AA\ (log$_{10}T \rm{[K]}=6.2$) and 211\,\AA\ (log$_{10}T \rm{[K]}=6.3$) filters of AIA, where numbers in parentheses correspond to the temperatures where the thermal responses of the respective channels peak. These level-1 AIA EUV images have a cadence of 12\,s and an image scale of 0.6\arcsec\,pixel$^{-1}$. We deconvolved the AIA images with the point spread functions (PSF) of the respective channels using the \texttt{AIAPY} Python package \citep[][]{Barnes2020,Barnes2021}. This step helps in the reduction of artifacts in the images caused by PSF diffraction and filter mesh. The SDO HMI and AIA data are all processed using the \texttt{aia\_prep} procedure available in the solarsoft library \citep[][]{1998SoPh..182..497F}. The resulting images have an image scale of 0.6\arcsec\,pixel$^{-1}$ with solar north oriented up.

The Sun was quiescent with no visible active regions on the solar disk. The target was a network magnetic patch located in the north western quadrant, away from the disk center (Fig.\,\ref{fig:over}a--b). The MiHI field of view covered the positive polarity of the network patch and in the continuum images, these magnetic concentrations appear as typical clusters of bright points embedded within the granulation (Fig.\,\ref{fig:over}c). 

To further examine the dynamics of these magnetic patches, we have constructed blue-wing and red-wing images by averaging the H$\alpha$ line profile, at every pixel, in a wavelength window of $\pm0.15$\,\AA\ placed at $-1$\,\AA\ (for blue-wing) and $+1$\,\AA\ (for red-wing) from its nominal core position at 6562.817\,\AA. One such red-wing map from the time series is displayed in Fig.\,\ref{fig:over}d. The magnetic network bright point concentrations are now seen with higher intensity contrast with respect to the surrounding granulation as expected due to the hot-wall effect \citep[e.g.,][]{1976SoPh...50..269S,2006A&A...449.1209L}. 

We created integrated Stokes $V/I$ maps from the MiHI spectropolarimetric data to visualize the magnetic network patches. To improve the signal to noise ratio (S/N), we first averaged every 18 data sets to create Stokes $V/I$ cubes with a reduced cadence of 24\,s. Then the integration of the profile is done over the wavelength range of 0.5--0.9\,\AA\ on both the blue- and red-wings of the spectrum with respect to the nominal line core (Fig.\,\ref{fig:over}e). The unipolar positive character of the magnetic network patch is quite evident in the $V/I$ map.

We noticed that the H$\alpha$ wing images further reveal even higher intense compact brighter patches, that we refer to as hot spots, within the already bright network elements. One such example is highlighted by a red contour in Fig.\,\ref{fig:over}d. The descriptive term ``hot spot" has been used in the literature to refer to Ellerman bombs and photospheric bright points \citep[][]{2014A&A...567A.110B, 2017ApJS..229....5D}. In the following Section, we describe the detection of these hot spots and their properties. 

\begin{figure}
\begin{center}
\includegraphics[width=0.47\textwidth]{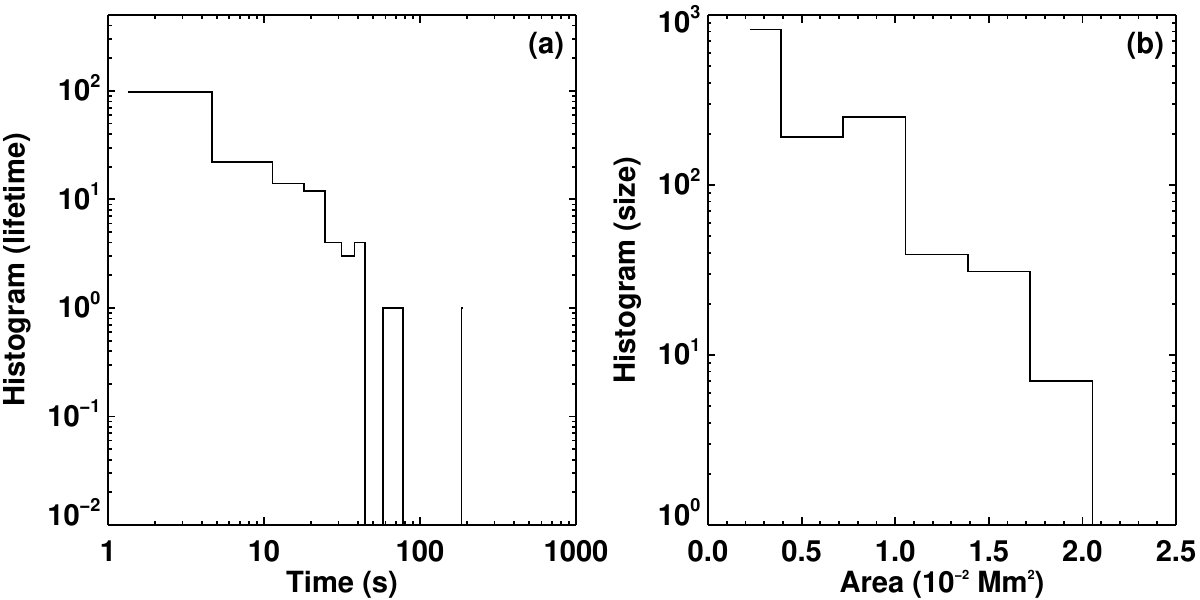}
\caption{Statistical properties of hot spots. Histograms of lifetimes detected events (panel a) and the size of individual events at every instance throughout their lifetime (panel b) are shown. See Sect.\,\ref{sec:det} for details. \label{fig:hist}}
\end{center}
\end{figure}

\section{Detection and Properties of photospheric hot spots\label{sec:det}}

The hot spots in both the blue and red-wing H$\alpha$ images are seen as compact patches with intensities above the ambient quiet-Sun at those wavelengths. We used this criterion to detect the hot spots in an automated way. To this end, we identified a quiet-Sun region outside of the network concentrations (2.67\arcsec$\times$1.37\arcsec\ sized magenta box overlaid on Fig.\,\ref{fig:events1}a). We first computed a mean H$\alpha$ map as a function of wavelength by temporally averaging the data for a period of 4.2\,minutes (UT\,09:26:53 to UT\,09:31:06). We then spatially averaged the intensity from the boxed region. The resulting average quiet-Sun H$\alpha$ line profile is plotted in Fig.\,\ref{fig:events1}b (magenta curve). A typical line profile from a network element is also shown for comparison (green curve in Fig.\,\ref{fig:events1}b). The H$\alpha$ wings of the network element have higher intensities compared to the quiet-Sun profile. Hot spots, like those from Fig.\,\ref{fig:over}d, have even higher wing intensities compared to the network element. Based on this, we define hot spots as those spatial pixels that exhibit enhancements simultaneously in the blue- and red-wing maps (wavelength window defined in Sect.\,\ref{sec:obs}) that are at least 1.4$\times$ the respective average blue- and red-wing intensity from the mean quiet-Sun H$\alpha$ spectrum.

We flagged pixels that satisfy the above intensity threshold, at each time step, with the aim to spatially $(x,y)$ and temporally $(t)$ group and label connected spots. We used the \texttt{label\_region} procedure available in the Interactive Data Language (IDL) for this purpose. To be able to detect hot spots also in the very first and very last of the snapshots, we padded the time series with zero frames on both ends. We also applied the three-dimensional (3D) \texttt{morph\_close} operator of IDL with kernel $(x,y,t)=(3,3,3)$ to fill any spatial and temporal holes (limited to a gap of 3 units) in our initial flagging. These steps yield for a detection of 161 hot spots, in total, in our 66\,minutes of MiHI observations (see Appendix\,\ref{app:detect} for further discussion on the detection methodology).

\subsection{Statistical Properties\label{sec:stat}}

Examples from the detected hot spots are displayed in Figures\,\ref{fig:events1} and \ref{fig:events2}. The hot spots clearly stand out above the quiet-Sun and network intensities. For each of the hot spots, we calculated its lifetime, and also the area at every instance that we could trace it. The histograms of these quantities are displayed in Fig.\,\ref{fig:hist}. The hot spots have lifetimes, generally below 100\,s, with the detectability of the shortest living ones limited by the cadence of the data (1.33\,s), with no clear peak in the distribution. Their area is less than one-hundredth of the area of granule (side length of 1\,Mm), with smaller ones having sizes limited by the image scale. Going back to Fig.\,\ref{fig:events1}, hot spots $\#1$ and $\#133$ are longer-living features, each lasting for about 30.67\,s and 22.67\,s, respectively. In Fig.\,\ref{fig:events2} we show two events that were detectable in only a single snapshot. 

Overall, the temporal filling factor, defined as the fractional duration of MiHI observations with at least one hot spot in a given frame, $f_t\approx0.36$, with an average lifetime of 11.14\,s. The average area of the hot spots is about 0.55$\times10^{-2}$\,Mm$^{2}$. Given that the hot spots are apparently closer to and within the magnetic network in our observations, we define the area filling factor of the hot spots with respect to the area coverage of the magnetic network. Throughout the observations the area occupied by the magnetic network elements has not considerably changed. Therefore, for simplicity, the area of magnetic network is determined by considering those pixels covered by cyan contours in Fig.\,\ref{fig:over}e. This yields, an area filling factor, defined as the ratio of average area of the hot spots to the area of the magnetic network, $f_a\approx10^{-3}$ (or 0.1\%).

\begin{figure}
\begin{center}
\includegraphics[width=0.47\textwidth]{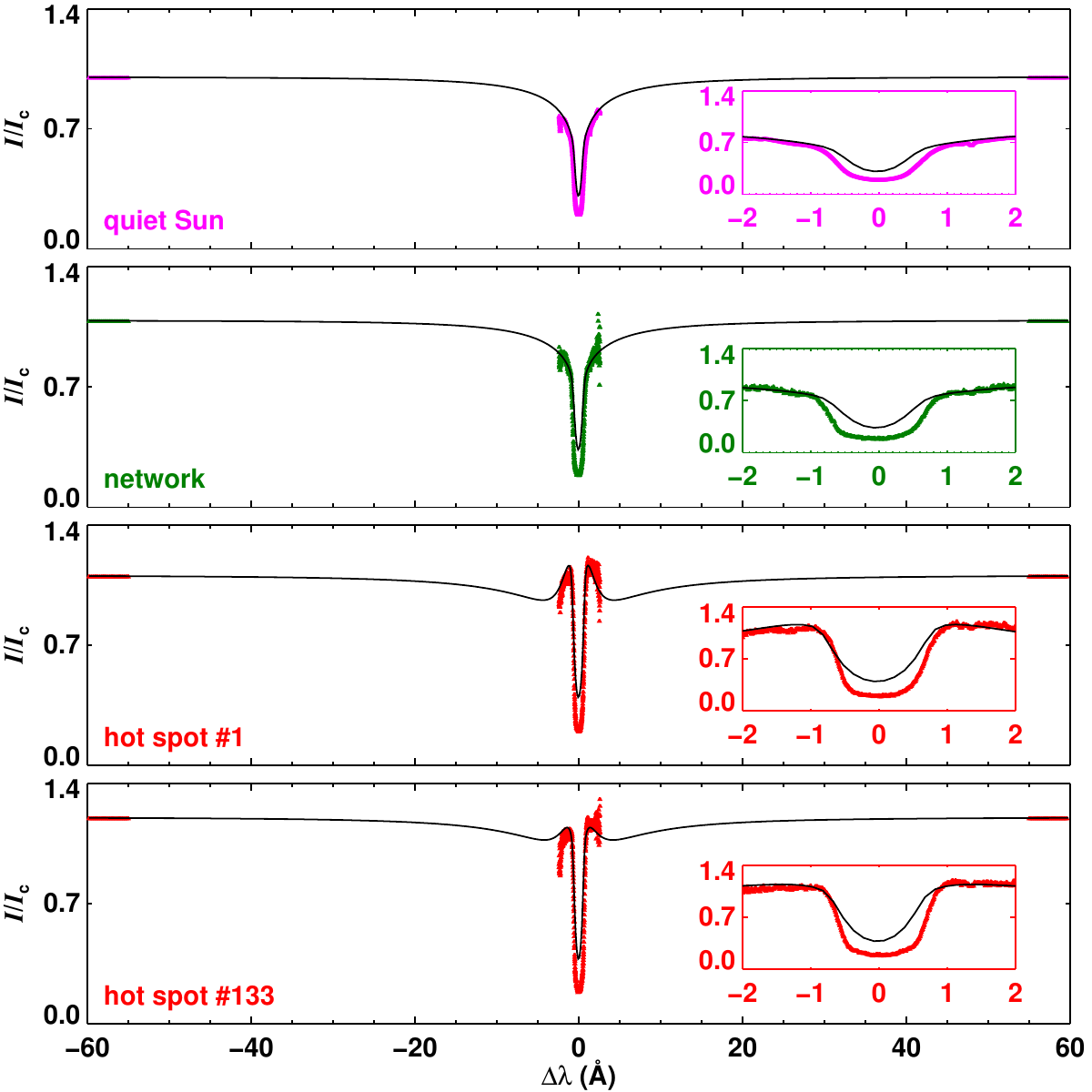}
\caption{Observed and inverted H$\alpha$ line profiles. The symbols around $\Delta\lambda=0$\,\AA\ in each panel are the observed H$\alpha$ Stokes\,$I$ profiles, normalized to the quiet-Sun continuum, of four different features as labeled. Continuum points are assumed to be on both blue- and red-wings of the observed line profile between 55\,\AA\ and 60\,\AA\ away from the nominal line core position (i.e., 6562.817\,\AA). We note that the observed continuum point is actually around 6468\,\AA\ (i.e., about 95\,\AA\ away from the line core on the blue-wing side). The observed  H$\alpha$ Stokes\,$I$ profiles including the assumed continuum points are normalized to the observed local continuum and further scaled with respect to the quiet-Sun continuum intensity. The solid black curves represent the best-fit inverted H$\alpha$ Stokes\,$I$ profiles. Insets in each panel zoom into the $\pm2$\,\AA\ about the core region of the respective H$\alpha$ profiles. See Sect.\,\ref{sec:ther} for details. \label{fig:prof}}
\end{center}
\end{figure}

\begin{figure}
\begin{center}
\includegraphics[width=0.47\textwidth]{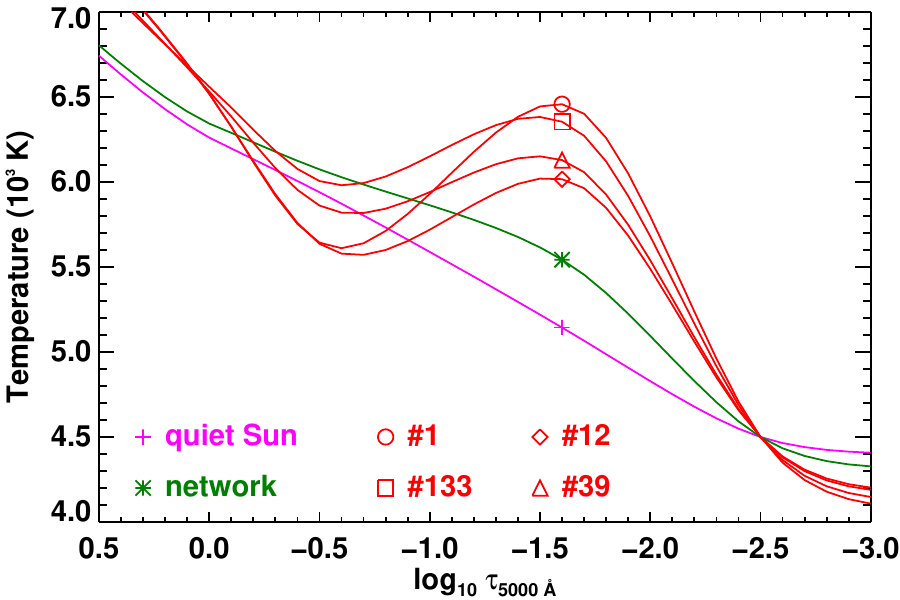}
\caption{Thermal diagnostics. Temperature stratification on optical depth scale of the inverted atmospheres corresponding to the various analyzed features as labeled. Four selected hot spot atmospheres are shown in red and are labeled by their event number. See Sect.\,\ref{sec:ther} for details.\label{fig:tempr}}
\end{center}
\end{figure}

\subsection{Thermal Properties\label{sec:ther}}
Clear enhancements in the wings of H$\alpha$ line profiles compared to the mean quiet-Sun and even magnetic network elements imply that the hot spots have different thermal/radiation characteristics. It is generally understood that the core of the H$\alpha$ line forms in the upper chromosphere under non local thermodynamic equilibrium (non-LTE) conditions, with non-equilibrium ionization of hydrogen playing an important role in the H$\alpha$ line opacity \citep[][]{2002ApJ...572..626C,2012ApJ...749..136L}. In the photosphere and low chromosphere, level populations are close to LTE and the  H$\alpha$ line opacity is more sensitive to the temperature \citep[][]{2012ApJ...749..136L}. Forward modeling suggests that the wings of the H$\alpha$ line could be considered to form under near-LTE conditions \citep[][]{2006A&A...449.1209L} and non-LTE departures are less than 1\% \citep[][]{2013A&A...554A.118P}.  

With this background, to decipher the nature of the observed H$\alpha$ wing enhancements in the hot spots, we performed a 1D spectral inversion of the H$\alpha$ line profile, including the continuum, with the LTE approximation for the line formation. To this end, we used SPINOR, which is a node-based LTE inversion code \citep[][]{2000A&A...358.1109F}. The recorded continuum wavelength point is about 95\,\AA\ blue-ward from the H$\alpha$ line core. This means that the SPINOR code has to work on a wavelength grid of nearly $\pm$95\,\AA\ (blue- and red-wing combined), where in actuality we have data coverage over only 5\,\AA\ around the H$\alpha$ line and nothing till the continuum point. To reduce the size of the spectral grid used by the inversion scheme and at the same time to maintain the very broad wing-to-continuum transition of the H$\alpha$ line, we moved the continuum points between 55\,\AA\ and 60\,\AA\ away from the line core on either side of the spectrum.

Given that our aim is to understand the thermal properties of the hot spots, we used a simple atmospheric model with four nodes for the temperature (at optical depth, $\log \tau_{5000}$ of $0,-0.5,-1.7,-2.5$), and one node each for the line of sight velocity and microturbulent velocity. Moreover, after experimenting with inversions, we determined to fix the temperature to 4500\,K at the highest node  ($\log \tau = -2.5$). This temperature is sufficient to well populate the $n=2$ level of hydrogen but not large enough to produce an emission in the line core. During the inversions, only the line wings and the continuum were fitted. The information in the line core was not used for the $\chi^2$ computations.

Inverted spectral profiles of various features are displayed in Fig.\,\ref{fig:prof}. Obviously, an LTE approximation is not applicable to the H$\alpha$ line core, but the blue- and red-wings, and the monotonic transition to the continuum seen in the quiet-Sun and network element spectral profiles are reproduced well. Interestingly, the inversions are able to reproduce the enhanced wings seen in the hot spots. This is associated with a clear difference in the thermal structuring of various features. The quiet-Sun and network element show a monotonic decrease in the temperature with decreasing $\log \tau$, whereas the hot spots show about 1000\,K enhancements over the quiet-Sun value around $\log \tau = -1.7$ (see Figure\,\ref{fig:tempr} and Appendix\,\ref{app:therm} for additional discussion on thermal properties of hot spots). In comparison, the network element has a temperature increment of about 500\,K over the quiet-Sun value at the same optical depth. These temperature enhancements in the hot spots  can be lower than what is inferred from the forward modeling of Ellerman bomb type H$\alpha$ profiles in active regions. An example of the latter is found in \citet{2014A&A...567A.110B}, where the authors impose temperature enhancements of a few 1000\,K in the height range of up to 1000\,km above the photosphere to simulate wing enhancements in active region Ellerman bomb type H$\alpha$ profiles. Thus, our results are generally consistent with the notion that as the magnetic pressure in magnetic concentrations reduces the density, radiation can escape from deeper, hotter layers than in the surrounding area \citep[][]{1976SoPh...50..269S}. 

\section{Discussion\label{sec:disc}}

Compared to the profiles from the network elements, hot spots exhibit a  mustache-like wing enhancement that is generally observed in Ellerman bombs \citep[][]{1917ApJ....46..298E}. In particular, their wing enhancement excess and lifetimes are comparable to the Ellerman bombs observed in quiet-Sun regions \citep[][]{2016A&A...592A.100R}. Ellerman bombs are triggered by magnetic reconnection during flux emergence and cancellation process \citep[e.g.,][]{2002ApJ...575..506G,2007ApJ...657L..53I,2013ApJ...779..125N,2017A&A...601A.122D,2017ApJ...839...22H}. Internetwork magnetic elements frequently cancel around the magnetic network \citep[][]{2014ApJ...797...49G}. Therefore it is possible that the occurrence of these hot spots could be spatially associated with magnetic reconnection during cancellation of weak and rapidly evolving internetwork magnetic elements closer to the magnetic network \citep[][]{2017A&A...598A..47A,2017ApJS..229...17S,2018ApJ...862L..24P,2019A&A...623A.176C,2023ApJ...956L...1C} and in turn to the Ellerman bombs \citep[][]{2016A&A...592A.100R,2022A&A...664A..72J}. In fact, lifetimes and areas of hots spots determined here are comparable to the properties of some of the quiet-Sun Ellerman bombs \citep[][]{2020A&A...641L...5J,2022A&A...664A..72J}.

In this data set, the Stokes $V/I$ maps do not show any convincing evidence for the existence of the opposite (negative) polarity magnetic field patches, at any point in time, that are required to trigger the reconnection to power Ellerman bombs (see movie associated with Fig.\,\ref{fig:over}e). Given that the H$\alpha$ line is not particularly sensitive to magnetic fields, it could still be that there are hidden parasitic magnetic patches \citep[e.g.,][]{2016ApJ...820L..13W}, below the sensitivity limits of the observations that are creating the hot spots by locally heating the plasma through reconnection triggered by flux cancellation. Indeed, \citet{2020A&A...641L...5J} noted that some of the quiet-Sun Ellerman bombs may appear in unipolar magnetic patches, which prompts the question whether this sub-class of events reported in the literature is also driven by magnetic reconnection. Could reconnection due to small misalignments in the magnetic field, a process invoked to explain coronal heating \citep[][]{1988ApJ...330..474P}, also be operating in the photosphere and heating hot spots? The viability of such a process under photospheric high plasma-beta conditions is not obvious.

However, an alternative interpretation of the hot spots that does not rely on the existence of hidden opposite polarity magnetic patches is plausible. We found that the hot spot events display a stronger asymmetry towards the red wing than the mean quiet-Sun and network profiles (Appendix\,\ref{app:bisec} and Fig.\,\ref{fig:bisect}). This red-wing asymmetry indicates (convection-driven) downdrafts at these locations. Additionally, hot spots also exhibit enhanced circular polarization signal than the surrounding network elements (Appendix\,\ref{app:magsig} and Fig.\,\ref{fig:hsp}). Based on these two effects, we suggest that the hot spots are regions of enhanced magnetic field strength. 

MHD simulations reveal that magnetic field strength can be enhanced by compression due to converging granules acting on the magnetic elements in intergranular lanes, on timescales as short as 30\,s, to longer timescales of 100\,s \citep[][]{2020A&A...633A..60K}. The lifetime of this process based on MHD simulations is somewhat comparable to the average lifetime of hot spots ($\sim$11\,s). This would then imply that the strong wing enhancements in the hot spots are a consequence of radiation escaping from the deeper, hotter layers during phases of enhanced magnetic field strength. Bright patches of potentially similar nature to those studied here were also found in the cores of Fe\,{\sc i} 6297.79\,\AA, 6301.49\,\AA, and 6302.50\,\AA\ lines in facular regions hosting stronger magnetic fields \citep[][]{Saranathan2021}.  

Unlike Ellerman bombs that are thought to be imprints of local plasma heating due to magnetic reconnection, there is no additional heating occurring at the site of hot spots in our interpretation. This view aligns with the arguments of \citet{2013JPhCS.440a2007R}, who argued that network bright points associated with strong magnetic field concentrations were misinterpreted as Ellerman bombs in some of earlier studies.

\begin{figure}
\begin{center}
\includegraphics[width=0.47\textwidth]{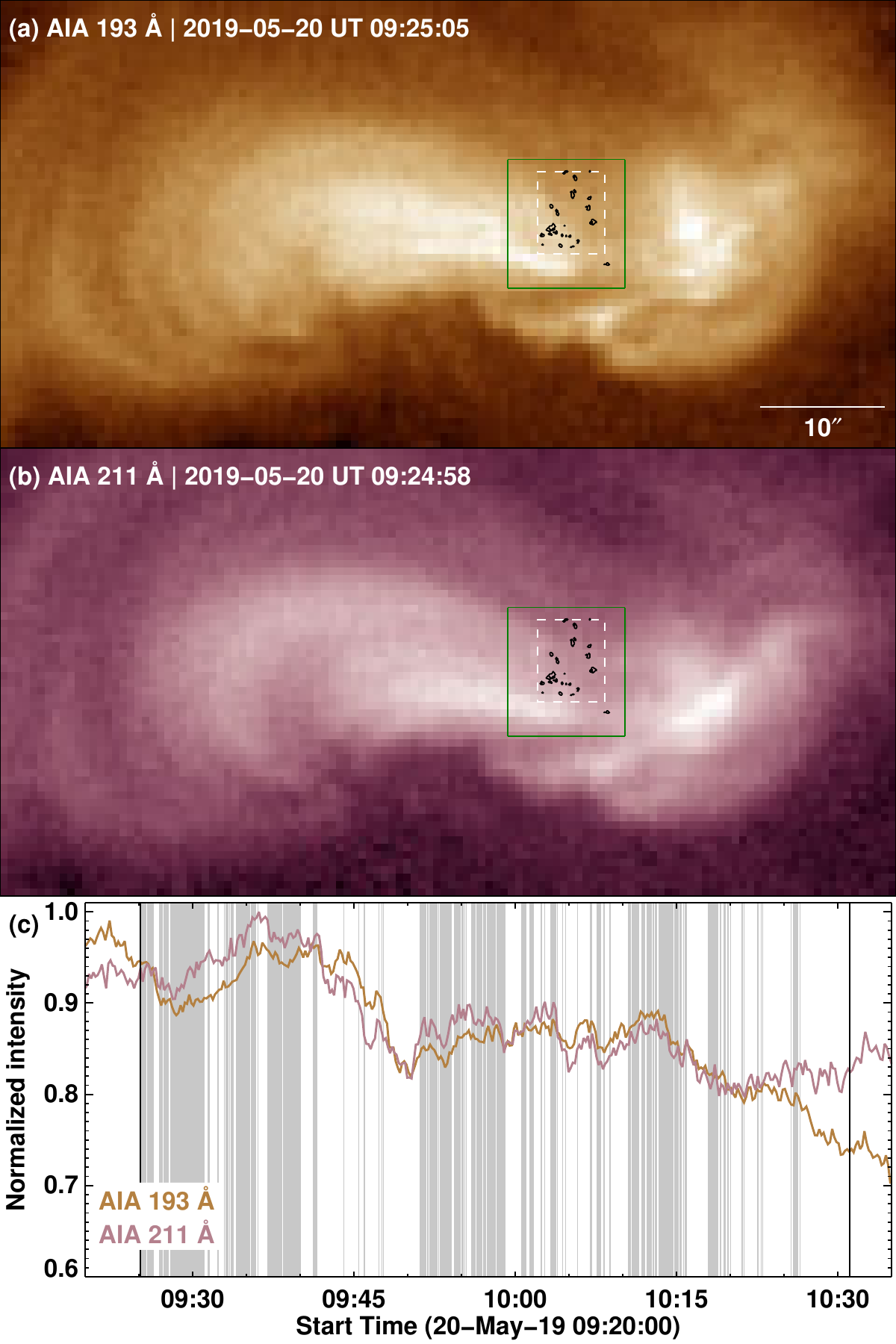}
\caption{Coronal evolution. Panels a and b show the coronal EUV emission maps recorded by the AIA 193\,\AA\ and 211\,\AA\ filters. The green box outlines the MiHI FOV (see Fig.\,\ref{fig:over}). The white dashed box marks the area over which we detected hot spots in our MiHI H$\alpha$ data. The black contours represent the locations where these hot spots were detected over the course of $\sim$65\,minutes of observations. In panel c we plot the EUV emission light curves from the two filters (brown and purple colored curves). These light curves are mean EUV emission from the respective white dashed box. The gray vertical lines mark the instances when we detected hot spots. In this case, a vertical line is drawn if a given MiHI snapshot has at least one hot spot in the FOV. See Sects.\,\ref{sec:obs} and \ref{sec:disc} for details.\label{fig:lc}}
\end{center}
\end{figure}

The MiHI hot spots are observed to occur directly at the base of million Kelvin coronal loops, as observed with SDO/AIA (black contours in Fig.\,\ref{fig:lc}a--b). This is interesting because our observations reveal that the magnetic field intensification process is a continual and ongoing one at the base of the coronal loops. Increasing magnetic field strength coupled with photospheric flows will channel Poynting flux into the solar atmosphere.

Here we provide an order of magnitude estimate of the energy flux transfer associated with the convection-driven magnetic field intensification. During the field intensification, increased downflows (signatures seen as red wing asymmetry), may lead to the compression of the flux tube. Here we assume that the flux tube is compressed to a smaller size of 50\,km\ from an initial radius of 100\,km. We found the average lifetime of hot spots is $\sim$11\,s. This means that the field lines are dragged inward by 50\,km during this time with a velocity of 4.5\,km\,s$^{-1}$, which is in the range of speeds of horizontal motions exhibited by magnetic elements in the quiet Sun \citep[e.g.,][]{2012ApJ...752...48C,2013A&A...549A.116J}. 

Magnetic flux conservation implies the field strength in the compressed state is a factor of 4 higher than in the initial configuration. Indeed, hot spots appear to carry a stronger polarization signal than the surroundings (Appendix\,\ref{app:magsig}). This stronger signal could be a result of hot spots sampling deeper layers and thus at a given optical depth, the local field strengths are stronger than the surroundings network elements.

The average vertical component of the Poynting flux, $\left<S_z\right> = v_hB_hB_zf_tf_a/4\pi$, where $v_h=4.5$\,km\,s$^{-1}$ is the horizontal velocity, and $B_z=1.6$\,kG is the strength of the compressed flux tube \citep[field strength comparable to that of a network element][]{2019A&A...630A..86B}. Any convective downdrafts along $B_z$ will not contribute to the Poynting flux. For the horizontal magnetic field strength we assume $B_h\approx50$\,G, the value determined for quiet-Sun internetwork field \citep[][]{2008ApJ...672.1237L}, a reasonable assumption considering the region of interest in our study. $f_t\approx0.36$ and $f_a\approx10^{-3}$ are the temporal and area filling factor of hots spots (c.f. Sect.\,\ref{sec:stat}). With all these values, we determine $\left<S_z\right> =10^{6}$\,erg\,cm$^{-2}$\,s$^{-1}$. This estimate falls within the range of vertical components of the Poynting flux estimated in the quiet-Sun regions \citep[e.g.,][]{2023ApJ...956...83T}. 

Despite their low area filling factor, the relatively larger temporal filling factor means that hot spots and the underlying convection-driven field intensification (caused by granular compression), is a common phenomenon in the photosphere. But our estimate of the average vertical Poynting flux corresponds only to the intensification phase. Once the network element is properly established with field strengths exceeding 1\,kG, it can continue to inject energy flux into the upper atmosphere through different processes including magnetic braiding \citep[][]{1972ApJ...174..499P}, magnetohydrodynamic waves \citep[][]{2020SSRv..216..140V} or footpoint reconnection \citep[][]{2018ApJ...862L..24P}. It is interesting to note that the value of the Poynting flux we obtained is of the same order of magnitude as the average quiet-Sun coronal energy losses \citep[][]{1977ARA&A..15..363W}. The occurrence of hot spot features with respect to the coronal emission is shown in Fig.\,\ref{fig:lc}c. Here we are not suggesting that hot spots themselves are directly related to the coronal emission, but they might serve as proxies for photospheric locations of energy injection into the corona during the initial phases of magnetic field intensification. The chromosphere in our observations is, as expected, quite dynamic. How this energy flux generated through field amplification is transferred upward and whether it leads to any clear signatures separate from other processes (e.g., waves) in the chromosphere, requires further investigation.

\section{Conclusions\label{sec:conc}}

Using MiHI hyperspectral observations, we observed persistent H$\alpha$ wing-enhancement events in a magnetic network patch at the footpoints of coronal loops in the quiet-Sun region (Figs.\,\ref{fig:over}, \ref{fig:lc}). These hot spots appear brighter in the H$\alpha$ wings than the typical magnetic network elements, implying a clear difference in their thermal structuring compared to the surrounding regions (Figs.\,\ref{fig:events1}, \ref{fig:events2}, \ref{fig:prof}, \ref{fig:tempr}). Their average area of about 0.55$\times10^{-2}$\,Mm$^{2}$ and average lifetime of 11.14\,s (Fig.\,\ref{fig:hist}), point to a dynamic evolution. The line profiles from the hot spots resemble those from reconnection-driven Ellerman bombs, but it is unclear whether this is the correct explanation in view of the absence here of high-quality magnetograms that could unambiguously establish the lack of opposite polarity magnetic field patches at these locations. 

However, their pronounced red-wing asymmetries (Fig.\,\ref{fig:bisect}) suggest that the hot spots may instead be related to convection-driven magnetic field intensification in the photosphere. In long-lived magnetic concentrations, repeated granular interactions could lead to distinct events at the same location. Temporally disconnected events occurring at the same location can be identified in the movie associated with Fig.\,\ref{fig:events1}. Magnetic field intensification will lead to enhanced Poynting flux transfer to higher layers.  We suggest that these photospheric hot spots at the base of coronal loops may therefore mark spatial and temporal locations for enhanced energy injection into the solar atmosphere.

%
%-------------------------------------------------------------------
%

\begin{acknowledgments}
We thank the two anonymous referees for constructive comments that helped us improve the manuscript. The authors thank S. K. Solanki and C. J. D\'iaz Baso for helpful discussions. L.P.C. gratefully acknowledges funding by the European Union (ERC, ORIGIN, 101039844). Views and opinions expressed are however those of the author(s) only and do not necessarily reflect those of the European Union or the European Research Council. Neither the European Union nor the granting authority can be held responsible for them. L.R.V.D.V. is supported by the Research Council of Norway, project number 325491, and through its Centres of Excellence scheme, project number 262622. The Swedish 1-m Solar Telescope is operated on the island of La Palma by the Institute for Solar Physics of Stockholm University in the Spanish Observatorio del Roque de los Muchachos of the Instituto de Astrofísica de Canarias. The Institute for Solar Physics is supported by a grant for research infrastructures of national importance from the Swedish Research Council (registration number 2021-00169). SDO is the first mission to be launched for NASA's Living With a Star (LWS) Program and the data are supplied courtesy of the SDO/HMI and SDO/AIA consortia. This research was supported by the International Space Science Institute (ISSI) in Bern, through ISSI International Team project \#23-586 (Novel Insights Into Bursts, Bombs, and Brightenings in the Solar Atmosphere from Solar Orbiter). We acknowledge the usage of {\sc aiapy} open source software package \citep[v0.7.3;][]{Barnes2020,Barnes2021}, {\sc sunpy} \citep[][]{2020ApJ...890...68S}, {\sc astropy} \citep[][]{2022ApJ...935..167A}, {\sc sswidl} \citep[][]{1998SoPh..182..497F}.
\end{acknowledgments}

%
%-------------------------------------------------------------------
%

\appendix
\section{Discussion on detection methodology\label{app:detect}}
We described the detection methodology of hot spots in Sect.\,\ref{sec:det}. In summary, we first flag pixels exhibiting blue- and red-wing enhancements of at least 1.4$\times$ the quiet-Sun spectrum. We then use morphological operators to group the pixels into events. This method yielded a total of 161 hot spot events in our MiHI observations. The threshold value of 1.4 is chosen such that we detect only bright pixels that stand well above the network features. But this hard limit and pixel grouping could also lead to mis-identification of longer living features. In such a case, a longer living feature might be categorized into several features with shorter lifetimes. For instance, between the time range UT\,09:25:09 and UT\,09:41:32, we identified several hot spots from the labeled events \#1 through \#52, to occur within a distance of less than 0.5\arcsec\ from each other (see animation associated with Fig.\,\ref{fig:events1}). Could all or several of these events occurring at that location during that 16\,minute interval be identified as fewer or even a single longer-living event? It is not a very clear association because the corresponding bright point patches at that location show substantial evolution (including translation motions, merging and splitting) during the course of this 16 minutes interval. 

In any case, to be able to identify and group several of those events to a fewer longer-living ones, we would need to reduce our initial flagging threshold to between 1.35 and 1.3. If we set our new threshold to 1.3, then while the lifetimes of some of the existing hot spots become longer, we also detect other newer shorter living hot spots. Indeed, we found that there is a four-fold increase in the total number of hot spot events (from 161 events with threshold of 1.4 increase to 741 events with the new threshold of 1.3). The distribution of lifetimes is still skewed toward the lower end, with a new mean lifetime of 14\,s (instead of 11\,s as discussed in the main text). But the spatial and temporal filling factors, each increase by roughly two times. This would then lead to a four-fold increase in our Poynting flux estimates (see Sect.\,\ref{sec:disc}).

Another way to group the pixels would be to increase the kernel size used in the \texttt{morph\_close} operator. In the main text, we discussed results with a kernel size of 3 units (in both spatial and temporal directions) to group or label the pixels and found 161 events. But if we increase the kernel size to 5, 7, and 9 units then the number of events reduce to 125, 113, and 102, respectively. With increasing kernel sizes, more and more pixels with lower wing enhancements get grouped into events. Accordingly, the spatial and temporal filling factors increase.

Overall, our key result, namely, the existence of hot spots at the base of coronal loops, is not affected by how we group the pixels into ``events''. We further emphasize that the ``number of events'' does not enter our Poynting flux estimate. Moreover, at a given intensity threshold, both spatial and temporal filling factors can be considered lower limits because our definition of a hot spot is very strict, as we only consider strong cases that are well above the wing enhancements of network patches.

\section{Additional discussion on thermal properties\label{app:therm}}
In Fig.\,\ref{fig:tempr}, we presented the thermal characteristics of four hot spots. The main point is that the features exhibit temperature enhancements of about 1000\,K\ over the quiet-Sun values around $\log \tau = -1.7$. During inversions, we identified that the wing enhancements are most sensitive to the temperature at the node $\log \tau = -1.7$. To investigate the robustness of our results and to understand how random variations in temperature affect the inverted line profiles, we carried out a temperature sensitivity analysis. Unfortunately, the spectral response to temperature changes in several nodes is relatively similar, so that changes in one node can be partially compensated with changes in other nodes. To include this coupling in the temperature sensitivity analysis, we carried out a Monte-Carlo type analysis, where the temperatures of all nodes are perturbed according to a Gaussian probability distribution.

\begin{figure}
\begin{center}
\includegraphics[width=0.47\textwidth]{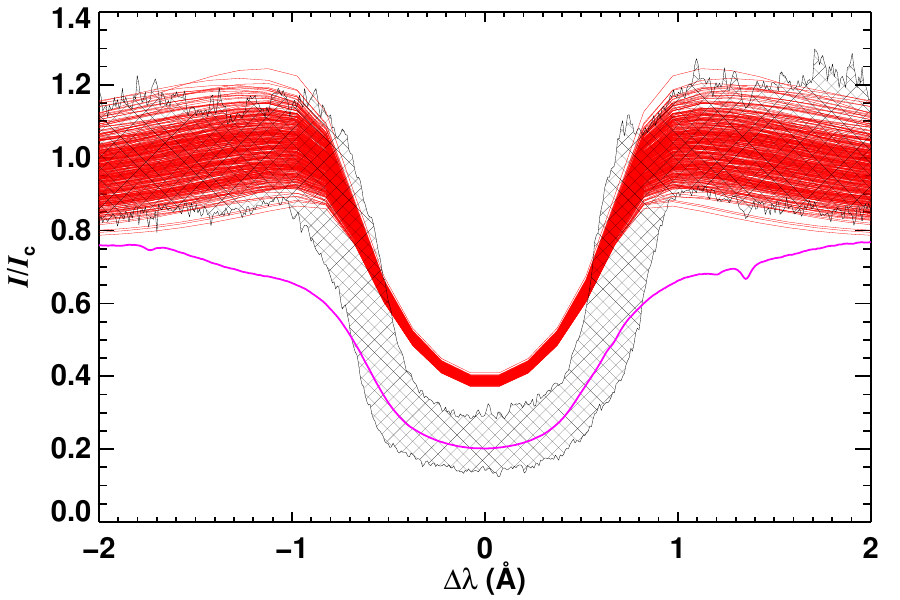}
\caption{Forward synthesis of H$\alpha$ hot spots profiles. Black hatched region covers the entire range of the intensity profiles of all the 161 hot spots, retrieved during their brightness peak. The red colored curves are the synthesized H$\alpha$ line profiles obtained by randomly varying the temperature of the input atmosphere. The magenta curve is the quiet-Sun H$\alpha$ line profile, shown here for reference. See Sect.\,\ref{sec:ther} and Appendix\,\ref{app:therm} for details.\label{fig:prof_all}}
\end{center}
\end{figure}

We started with a temperature profile that reproduces wing enhancement of the mean profile of all the 161 hot spots at their brightness peak. We then constructed 400 realizations of new atmospheres around this reference atmosphere by randomly varying the temperature such that the standard deviation in the temperature distribution is about 125\,K, at the bottom three node (fixing the temperature at the top node to 4500\,K). We then synthesized H$\alpha$ line profiles corresponding to these 400 atmospheres and compared them with the observed line profiles of all the 161 hot spots (red curves and black hatched region in Fig.\,\ref{fig:prof_all}). This temperature distribution clearly reproduces the entire range of hot spot profiles, particularly the wing enhancement well above the quiet-Sun profile. This intensity range of the observed wing enhancement has a 3$\sigma$ extent around the mean profile. Therefore, we conclude that the temperature distribution of the hot spots will have standard deviation of about 125\,K, with all events being several 100\,K hotter than the quiet-Sun temperature at optical depths of $\log \tau = -1.7$.

A formal analysis of fit errors \citep[for example, Eq.\,42 of][]{2016LRSP...13....4D}, is not meaningful in our case as we are only fitting a part of the profile (wings and continuum). Moreover, there are no observations between the wing and the continuum points. By taking into account the effect of observational noise, the above experiment will also serve as a temperature sensitivity analysis of individual hot spots. With strong observational constraints provided by the wing enhancements and the continuum level, we infer that the inverted temperature profiles of individual hot spots will have uncertainties even smaller than 125\,K.

\section{H$\alpha$ line profile characteristics\label{app:bisec}}

\begin{figure}
\begin{center}
\includegraphics[width=0.47\textwidth]{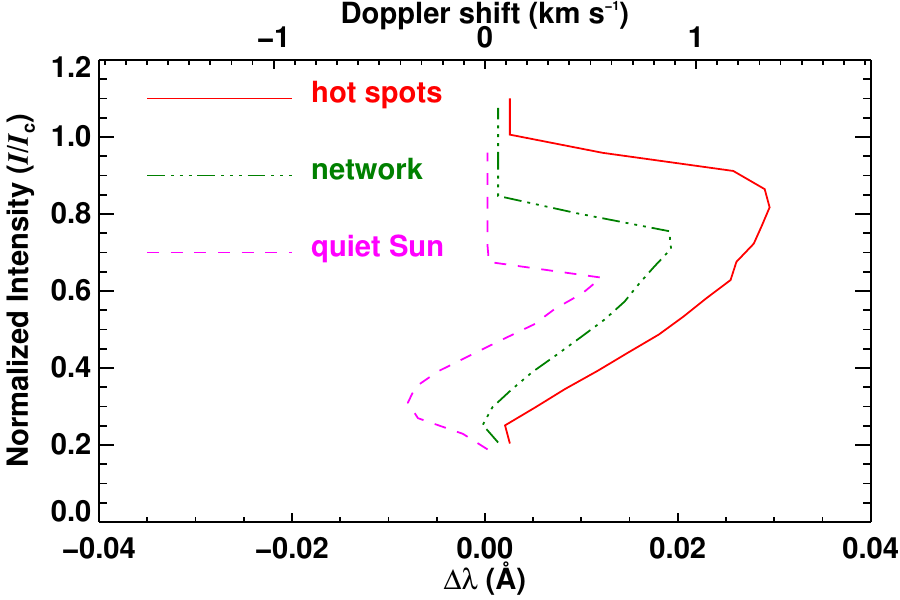}
\caption{H$\alpha$ line bisectors. The three curves show the line bisectors derived for the average H$\alpha$ line profile corresponding to the three distinct features as labeled (red: hot spots, green: network, magenta: quiet Sun). See Appendix\,\ref{app:bisec} for details. \label{fig:bisect}}
\end{center}
\end{figure}

We observed that the H$\alpha$ line profiles of hot spots show red wing asymmetry. To quantify this effect further, we have conducted line bisector analysis as follows. We first gathered all the hot spot profiles at spatial and temporal locations where they reach their intensity peaks. We then obtained a mean hot spot profile by averaging all the 161 profiles with respect to their line cores (i.e., we took into account any line shift of the profile when averaging). Similarly, we obtained the average quiet-Sun and network profiles. For the quite-Sun, we considered all the pixels covered by the magenta box in Fig.\,\ref{fig:events1} from 30 snapshots that are each 133\,s apart (basically covering the whole observing sequence). For the network case, we only considered those profiles,  from these 30 snapshots, that lie between intensity threshold levels of 1.15 and 1.3, with respect to the quiet-Sun red wing intensity (wavelength window is defined in Sect.\,\ref{sec:obs}). The spectral line bisectors thus computed are displayed in Fig.\,\ref{fig:bisect}.  Our results are consistent with those presented in \citet{2013SoPh..288...89C}. Additionally, our observations reveal stronger red-wing asymmetry in hot spots, which was not distinguished in the earlier study.

\section{Magnetic signal of hot spots\label{app:magsig}}
\begin{figure*}
\begin{center}
\includegraphics[width=0.8\textwidth]{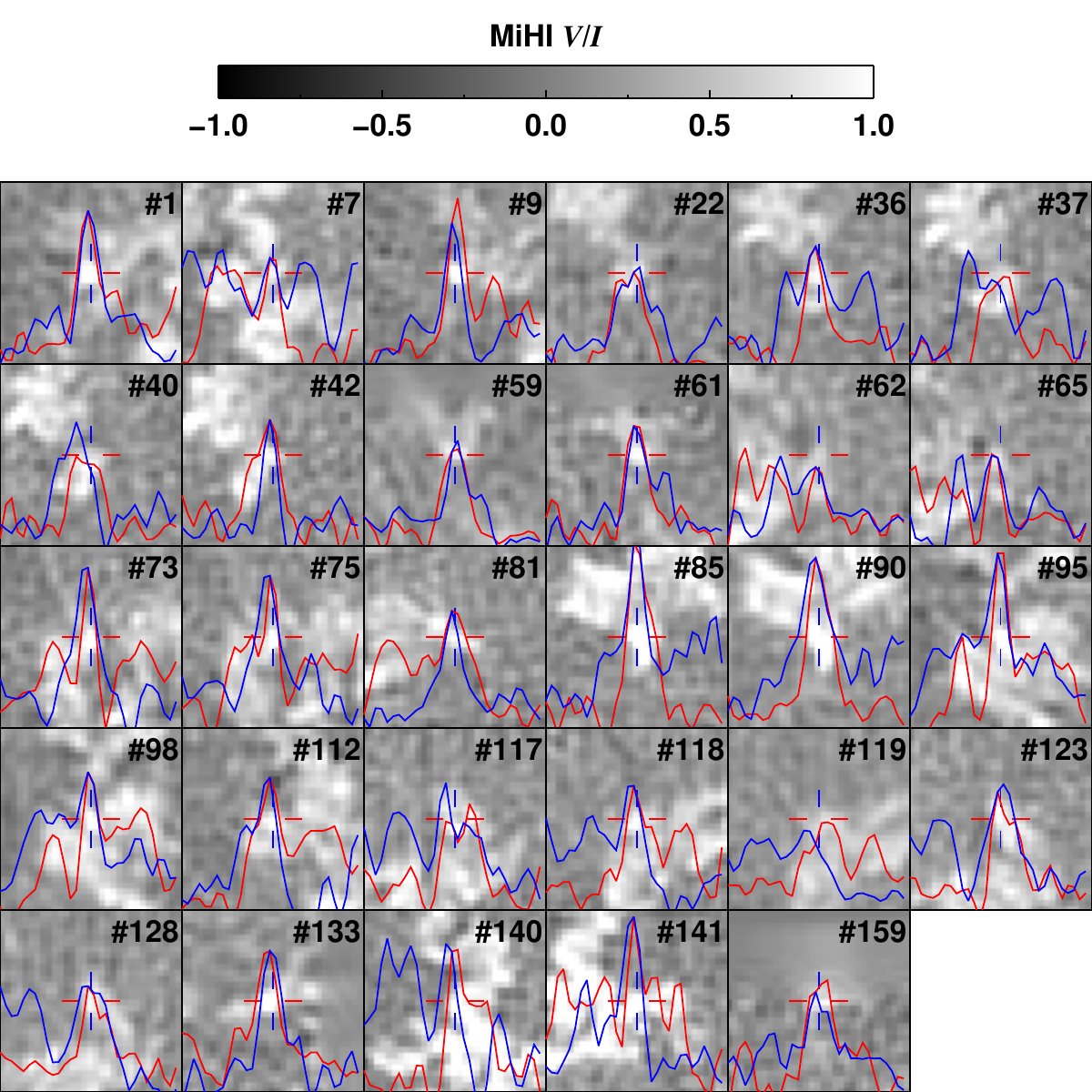}
\caption{Magnetic signal of hot spots. Integrated MiHI Stokes $V/I$ signal derived for the 29 hot spots whose lifetimes are at least 20\,s. The field of view (2\arcsec$\times$2\arcsec) is centered on the hot spot pixel, when the event reaches its peak intensity in the red wing (location marked by the red-blue cross-hair). The overlaid red and blue curves are the Stokes $V/I$ signal along the respective colored cross-hair section, over the extent of the field of view. All the line plots have the same scale, with minimum and maximum, respectively set to $0$ and $1.6$ in the Stokes $V/I$ units. See Sect.\,\ref{sec:obs} and Appendix\,\ref{app:magsig} for details. \label{fig:hsp}}
\end{center}
\end{figure*}

Integrated MiHI Stokes $V/I$ maps provide hints of enhanced polarization signal at the location of hot spots. This effect is noticeable in the animation associated with Fig.\,\ref{fig:over}e. To qualitatively investigate this effect further, we have identified hots spots whose lifetimes are at least 20\,s (i.e., comparable to the 24\,s cadence of the Stokes $V/I$ maps that we derived; see Sect.\,\ref{sec:obs}). There are 29 such hot spots in our sample with lifetimes of at least 20\,s. Then for each of these hot spots, we extracted a small patch of 2\arcsec$\times$2\arcsec\ area from the nearest-in-time Stokes $V/I$ map, centered around the brightest pixel of that particular event. All the 29 such patches are displayed in Fig.\,\ref{fig:hsp}, which shows concentrated signal at the center. The overlaid Stokes $V/I$ curves clearly show signal peak associated with the events. The signal is as strong or in some cases even stronger than the surrounding network elements.

\bibliographystyle{aasjournal}

\end{document}